\documentclass{nature}
\usepackage{graphicx}
\makeatletter
\let\saved@includegraphics\includegraphics
\AtBeginDocument{\let\includegraphics\saved@includegraphics}
\renewenvironment*{figure}{\@float{figure}}{\end@float}
\makeatother

\usepackage{multirow}
\usepackage{subfigure}
\usepackage{mathtools}
\usepackage[colorlinks=true,linkcolor=blue,citecolor=black,urlcolor=blue]{hyperref}
\usepackage{authblk}
\usepackage[labelfont=bf]{caption}
\usepackage[figurename=Figure]{caption}
\usepackage{siunitx}
\usepackage[dvipsnames]{xcolor}
\usepackage{soul}
\usepackage{xfrac}
\usepackage{amsmath}
\usepackage{soul,xcolor}
\usepackage{lineno}
\usepackage{siunitx}

\graphicspath{{./figures/}}

\setstcolor{red}

\title{A spin model for intrinsic antiferromagnetic skyrmions on  a triangular lattice}
\author{Amal Aldarawsheh$^{1,2*}$, Moritz Sallermann$^{1,3,4}$, Muayad Abusaa$^5$ and  Samir Lounis$^{1,2*}$}
\begin{document}
\maketitle
\begin{affiliations}
 \item Peter Gr\"{u}nberg Institute and Institute for Advanced Simulation, Forschungszentrum J\"{u}lich and JARA, D-52425 J\"{u}lich, Germany
 \item Faculty of Physics, University of Duisburg-Essen and CENIDE, 47053 Duisburg, Germany
 \item RWTH Aachen University, 52056 Aachen, Germany
 \item Science Institute and Faculty of Physical Sciences, University of Iceland, VR-III, 107 Reykjavík, Iceland
 \item Department of Physics, Arab American University, Jenin, Palestine\
 
 * a.aldarawsheh@fz-juelich.de; s.lounis@fz-juelich.de
 
\end{affiliations}
\section*{Abstract}

Skyrmions are prospected as the potential future of data storage due to their topologically protected spin structures. However, traditional ferromagnetic (FM) skyrmions experience deflection when  driven with an electric current, hindering their usage in spintronics. Antiferromagnetic (AFM) skyrmions, consisting of two FM solitons coupled antiferromagnetically, are predicted to have a zero Magnus force, making them promising candidates for spintronic racetrack memories. Currently, they have been stabilized in synthetic AFM structures, i.e. multilayers hosting FM skyrmions, which couple antiferromagnetically
through a non-magnetic spacer, while recent first-principles simulations predict their emergence in an intrinsic form,  within an row-wise AFM  single monolayer of Cr deposited on PdFe bilayer grown on Ir(111) surfaces. The latter material forms a triangular lattice, where single and interlinked AFM skyrmions can be stabilized. Here, we explore the minimal Heisenberg model enabling the occurrence of such AFM solitons and the underlying phase diagrams by accounting for the interplay between the Dzyaloshinskii-Moriya and  Heisenberg exchange interactions, as well as the magnetic anisotropy and impact of magnetic field. By providing the fundamental basis to identify and understand the behavior of intrinsic AFM skyrmions, we anticipate our model to become a powerful tool for exploring and designing new topological magnetic materials to conceptualize devices for AFM spintronics.

\textbf{Keywords:  Intrinsic antiferromagnetic skyrmions, spin model, single and interchained AFM skyrmions, triangular lattice, thermal stability, phase diagram, antiferromagnetism, topology.  }

\section*{Introduction}
Since their early observation~\cite{muhlbauer2009skyrmion,yu2010real,heinze2011spontaneous,romming2013writing}, skyrmions, which are magnetic textures with unique properties, have garnered the attention of the condensed matter community.  They are seen as potential bit representatives for future spintronic devices~\cite{}  due to their nontrivial topological twists and exotic properties~\cite{bogdanov1994thermodynamically,Roessler2006,kiselev2011chiral,Nagaosa2013,iwasaki2013universal,fert2013skyrmions,crum2015perpendicular,Fert2017,lai2017improved,fernandes2018universality,fernandes2020defect,zhang2020skyrmion, bogdanov2020physical,luo2021skyrmion,gobel2021beyond,lima2022spin,wang2022fundamental}. Skyrmion-based racetrack memory devices are expected to remarkably reduce the power consumption in data flow compared to domain walls~\cite{jonietz2010spin,yu2012skyrmion}.
However, ferromagnetic (FM) skyrmions are afflicted with  various drawbacks that limit their optimal utilization such as:  their sensitivity to stray fields and suffering from dipolar interactions~\cite{Woo2016,Buettner2018}, in addition to their complex response to applied currents leading to unwanted deflections~\cite{Nagaosa2013,Jiang2016}, which can become even more elaborated under the presence of defects~\cite{Lin2013,Jiang2016,Woo2016,fernandes2020impurity,reichhardt2022statics}. In contrast,  antiferromagnetic  (AFM) skyrmions have several advantages over their FM counterparts since the stray field cancels out~\cite{olejnik2018terahertz,WuzhangFang2021} augmented with an immunity to the Magnus force~\cite{barker2016static,zhang2016antiferromagnetic,velkov2016phenomenology,jin2016dynamics,gobel2017antiferromagnetic,akosa2018theory} with their potential  for ultrafast dynamics~\cite{gomonay2018antiferromagnetic}, and ability to overcome  defects~\cite{tomasello2017performance,silva2019antiferromagnetic}.

Several recent theoretical studies are inspecting the realization of individual AFM skyrmions or their periodic arrangement  assuming a squared~\cite{barker2016static,zhang2016antiferromagnetic,keesman2016skyrmions,jin2016dynamics,akosa2018theory}, a triangular~\cite{rosales2015three,mohylna2022spontaneous}  or a honeycomb lattice~\cite{gobel2017antiferromagnetic}. On the experimental side, synthetic AFM skyrmions were unveiled experimentally in multilayers, where FM films host regular FM skyrmions with an interfilm coupling of AFM nature through various spacers~\cite{dohi2019formation,legrand2020room,finco2021imaging,juge2022skyrmions,chen2022controllable}, while   fractional antiferromagnetic skyrmion
lattice was stabilized in MnSc$_{2}$S$_{4}$ ~\cite{gao2020fractional,rosales2022anisotropy} , and complex topological AFM objects were found in a bulk phase~\cite{jani2021antiferromagnetic}. In Ref.~\cite{aldarawsheh2022emergence}, we predicted the emergence of intrinsic single and interchained AFM skyrmions on a triangular lattice of row-wise AFM (RW-AFM) Cr layer deposited on PdFe/Ir(111). The latter substrate became over the last decade a perfect bed system for a plethora of  phenomena pertaining to FM skyrmions~\cite{romming2013writing,dupe2014tailoring,Simon2014,crum2015perpendicular,romming2015field,dos2016chirality,fernandes2018universality,fernandes2020defect,Arjana2020,bouhassoune2021friedel,lima2022spin}. 

 The goal of the current work is to introduce a Heisenberg model that incorporates the essential magnetic interactions required to produce AFM skyrmions on a triangular lattice. We perform atomistic spin simulations  on the basis of the Landau-Lifschitz-Gilbert (LLG) equations as implemented in  the Spirit code~\cite{muller2019spirit}. We  consider the interplay between the exchange interactions, Dzyaloshinskii-Moriya interactions (DMI),  the magnetic anisotropy and the impact of an external magnetic field to establish the phase diagrams of the intrinsic AFM skyrmions while inspecting their stability via simulations based on the geodesic nudged elastic band method (GNEB)~\cite{bessarab2015method,muller2018duplication,muller2019spirit}. Our model offers a robust approach to comprehend the behavior of AFM skyrmions in a triangular lattice with the aim of understanding the required ingredients for their stabilization and to create novel materials and devices for AFM spintronics.\\
\textbf{Methods}\\
 In our study, we consider a two dimensional  Heisenberg model on a triangular lattice, equipped with Heisenberg exchange coupling, DMI, the magnetic anisotropy energy (MAE), and Zeeman term. The energy functional reads as follows:

\begin{equation}
 H= -\sum\limits_{<i,j>} J_{ij}\;\textbf{S}_{i}\cdot \textbf{S}_{j}  - \sum\limits_{<i,j>}\textbf{D}_{ij}\cdot (\textbf{S}_{i}\times \textbf{S}_{j})- K\sum\limits_{i}  (S_i ^z)^2  - \sum\limits_{i} h_i S_i^z,  
\end{equation}
where $i$ and $j$ are site indices, each carrying a magnetic moment.  $\textbf{S}$ is the unit vector of the magnetic moment. $J$ is the Heisenberg exchange coupling strength, being negative for an AFM interaction while $\textbf{D}$ is the DMI  vector, and $K$ the magnetic anisotropy energy per atom favoring an out-of-plane orientation if positive. $h_i= \mu_i B$  describes the Zeeman coupling to the atomic spin moment $\mu$ at site $i$, assuming $\mu = 1~\mu_B$ and an out-of-plane field. 
To explore the magnetic properties and emerging complex states we utilize the Landau-Lifshitz-equation (LLG) as implemented in the Spirit code~\cite{muller2019spirit}. The simulations were carried out with 100$^2$, 200$^2$ and 300$^2$ sites assuming periodic  boundary conditions to model the extended two dimensional Heisenberg Hamiltonian  at zero Kelvin.
 
\section*{Results}

  
  \subsection{Phase diagrams} 
  
\begin{figure}[h!]
	\begin{center}
	\includegraphics[width=1\columnwidth,keepaspectratio]{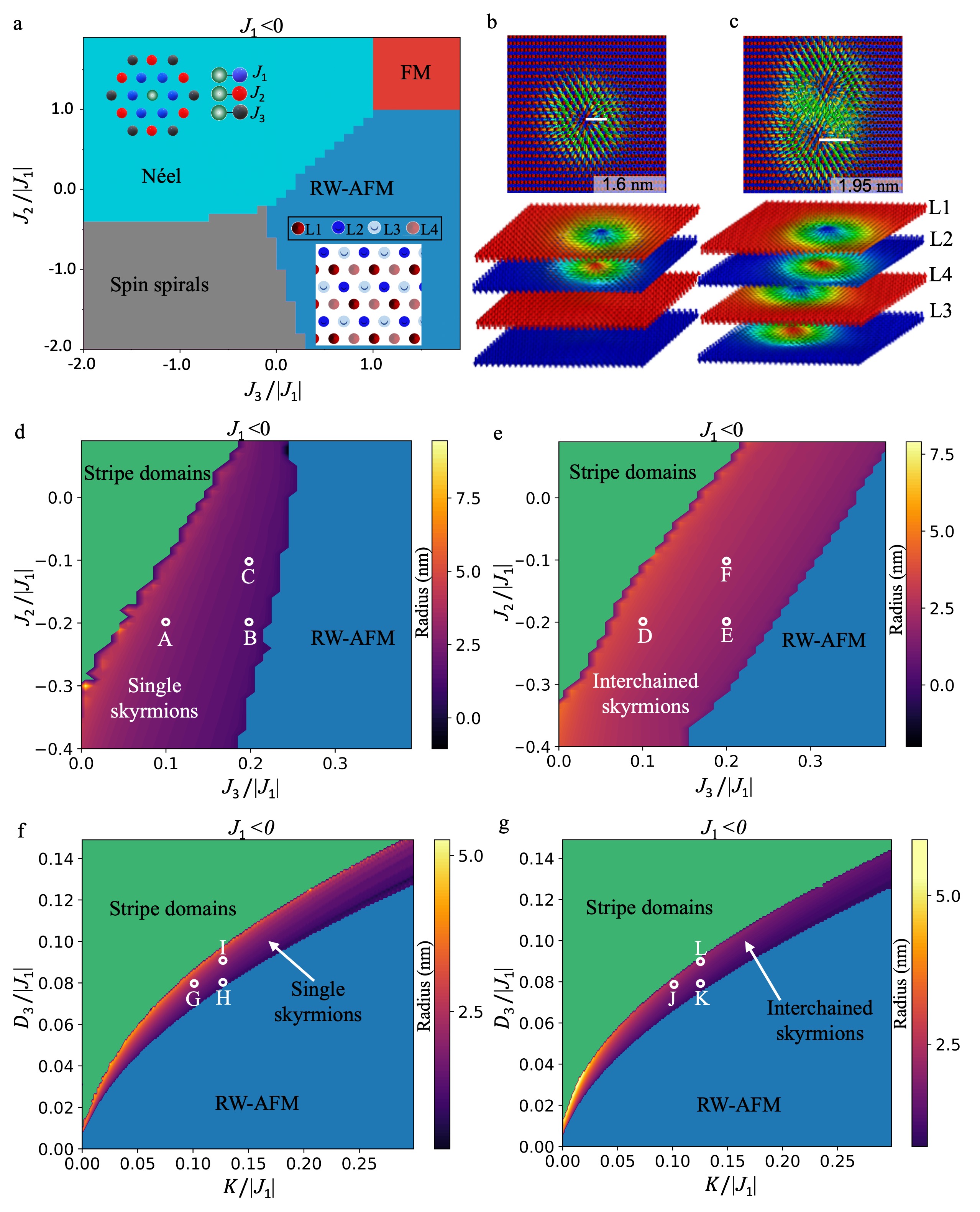} 
	\caption{\textbf{Phase diagrams for AFM skyrmions.}  \textbf{a} The 
	phase diagram for the Heisenberg model of a triangular lattice without spin-orbit induced interactions. The impact of magnetic interactions up to the third nearest neighbors is considered assuming an antiferromagnetic coupling $J_1$ among first nearest neighbors. The upper inset shows the triangular lattice indicating the considered magnetic exchange interactions while the lower inset illustrates the four sublattices building up the RW-AFM ground state.  \textbf{b} The single AFM skyrmion is made of two FM skyrmions, each hosted by sublattices L1 and L2, while L3 and L4 remain collinear, whereas the latter can host a second AFM skyrmion in case of the overlapped skyrmion scenario \textbf{c}. Note that the separation of the four sublattices in \textbf{b}-\textbf{c} is done for illustration purpose only since all skyrmions reside in the same layer. Phase diagram showing the range of interactions $J_{2}/|J_{1}|$ and $J_{3}/|J_{1}|$  at which the single \textbf{d} and double \textbf{e} AFM skyrmions can be stabilised with  $D_{3}/|J_{1}|$ =0.03, and $K/|J_{1}|$=0.024. The color code indicates the radius of the stabilized AFM skyrmion. \textbf{f} and \textbf{g} Phase diagrams obtained by changing  the  $D_{3}/|J_{1}|$ magnitude along with that of $K/|J_{1}|$ while fixing $J_{2}/|J_{1}|$ at -0.2, and  $J_{3}/|J_{1}|$ at 0.2, for single and double AFM skyrmions, respectively. The letters A-L shown in the diagrams indicate skyrmions which are plotted in Supplementary Figure S2.}   

		\label{fig:1}
	\end{center}
\end{figure}

 
 We start our investigations by determining the conditions to form a RW-AFM spin state, which  was observed experimentally in Mn/Re(0001)~\cite{spethmann2020discovery,spethmann2021discovery}. As established in Refs.~\cite{kurz2001non,hardrat2009complex}
 the minimum set of Heisenberg exchange interactions involves the interactions with first ($J_{1}$) and second ($J_2$) nearest neighboring interactions as  shown in the upper inset of Fig.~\ref{fig:1} a. The formation of the FM skyrmions building up our AFM skyrmions requires, as demonstrated in Ref.~\cite{aldarawsheh2022emergence}, a third nearest neighboring interaction  $J_3$, which should mediate a ferromagnetic coupling. Fig.~\ref{fig:1} a illustrates the underlying phase diagram, where we expect four regions that can host either a N\'eel, FM, spin spiraling and RW-AFM spin states. The dark blue color indicates the region of interest, where the magnetic moments are distributed into four sublattices L1, L2, L3 and L4 as shown in the lower inset of Fig.~\ref{fig:1} a.
 $J_3$ mediates the magnetic interaction within each sublattice and must be positive, i.e. favoring a FM alignment, to enable the stabilization of the RW-AFM state. If too weak with respect to $J_1$ or if it is of antiferromagnetic nature, either spin spirals or a N\'eel state are favored depending on the strength of $J_2$.
We observe that the RW-AFM configuration occupies a larger phase area  when $J_{2}$ is of AFM nature. 
 In the RW-AFM state, $J_3$ is thus positive, which together with the DMI vector $\textbf{D}_3$ that is connecting the third n.n. similarly to $J_3$, enable the formation of sublattice FM skyrmions.  $\textbf{D}_3$ lies in-plane and  is  perpendicular to the bond connecting  neighboring atoms as shown in Supplementary Figure S1. The AFM interaction among the FM skyrmions is mediated by $J_1$ such that the presence of $J_2$ is not requested. As predicted in Ref.~\cite{aldarawsheh2022emergence}, the single AFM skyrmion consists of FM skyrmions present in two sublattices (L1 and L2) with the other two sublattices remaining collinear, while for the double AFM skyrmions, the building blocks FM skyrmions reside in each of the four sublattices (L1, L2, L3 and L4) as illustrated in Figs.~\ref{fig:1}b-c.

After setting the base for the magnetic interactions needed to realize our AFM solitons, we inspect the range of parameters ($J_{2}$, $J_{3}$, $D_{3}$ and $K$) normalised to the absolute value of $J_{1}$,  within which the single and double interchained AFM skyrmions can be stabilized (Figs.~\ref{fig:1}d-g). The building blocks of the AFM solitons are FM skyrmions.  The region hosting the skyrmions, color coded in terms of their radius,  is sandwiched between the RW-AFM and stripe domains phases. Thus the impact of the underlying interactions is similar to what is expected from the FM topological objects. For instance increasing $J_3$ (Figs.~\ref{fig:1}d-e, with $K/|J_{1}|$ and $D_{3}/|J_{1}|$ equal to 0.024 and 0.03 respectively), which  defines the FM interaction among the spins of the FM skyrmions, or the magnetic anisotropy energy $K$ (Figs.~\ref{fig:1}f-g, with $J_2/|J_{1}|$ and $J_3/|J_{1}|$ equal to -0.2 and 0.2 respectively) shrinks the size of the spin-texture by ultimately leading to its annihilation, while the DM interaction $D_3$ induces the opposite behavior (Figs.~\ref{fig:1}f-g). Interestingly, $J_2$ counteracts $J_3$ by amplifying the skyrmion size,  which at some point can be deformed into stripe domains. For completeness, snapshots of skyrmions, labelled from A to L in Fig.~\ref{fig:1}, are presented in Supplementary Figure S2. 

It is significant to point out that the shape of the AFM skyrmions is determined by the specific interaction parameters involved. While the interactions between spins in one sublattice L$_{i}$, characterized by $J_{3}$ and $D_{3}$, lead to the formation of a FM skyrmion in that sublattice, the  interactions with spins in the other FM skyrmion hosting sublattice, governed by $J_{1}$ and $J_{2}$, have a substantial impact on shaping the resulting  AFM skyrmion, as illustrated in Supplementary Figure S3 and discussed in detail in Supplementary Note 1. Without $J_2$, there is an anisotropic cancellation of contributing magnetic interactions induced by the AFM alignment of the spins around the skyrmion, which triggers a shape elongation. A finite $J_2$ helps establishing a balance in the magnetic interaction, which reduces the aforementioned skyrmion asymmetry.  However, interlinking of the AFM skyrmions results in a slight deformation in the overlap region, as the spins in the overlapping area interact with the noncollinear spins of the second AFM skyrmion, which is not the case for the other spins on the free side of the AFM skyrmion.

It is worth mentioning that the size of the single AFM skyrmion is smaller than those participating in the formation of the interchained magnetic textures (see for example the radius given in Figs.~\ref{fig:1}b-c), which impacts on the details of the phase diagrams. On the one hand, the window in which the double AFM skyrmions are stabilised while varying $J_2$ and $J_3$ is larger than that of the single magnetic objects (Figs.\ref{fig:1}d-e). On the other hand, the single skyrmion phase seems wider and shifted to the upper region of the diagram while tuning $D_3$ and $K$.

\subsection{Response to external magnetic fields}
\begin{figure}[h!]
	\begin{center}
	\includegraphics[width=1\columnwidth,keepaspectratio]{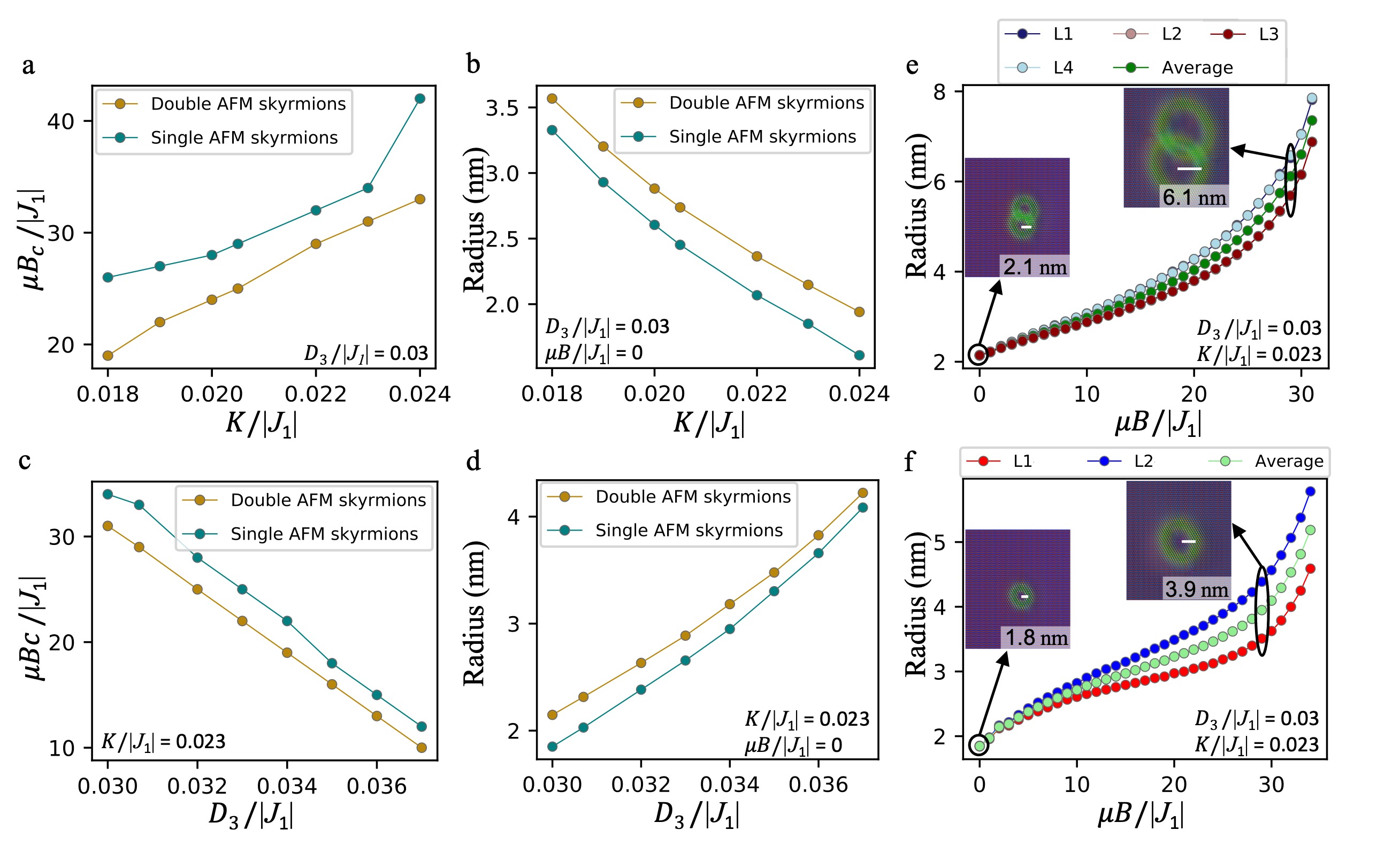} 
	\caption{\textbf{AFM skyrmions response to an external magnetic field.} 
	The critical magnetic field $\mu B_{c}/|J_{1}|$ tolerated by the single (green dots) and double (golden dots) AFM skyrmions as a function of \textbf{A} the normalized values of the anisotropy $K/|J_{1}|$, and \textbf{C} the normalized values of DMI $D_3/|J_1|$. \textbf{B} Increasing $K/|J_{1}|$ shrinks the radii of both the double and single AFM skyrmions, while \textbf{D} increasing $ D_3/|J_1|$ expands them. \textbf{E,F} Impact of the external magnetic field on the radii of both types of AFM skyrmions.}
		\label{fig:2}
	\end{center}
\end{figure}

The stability of skyrmions when exposed to an external magnetic field is an essential aspect for their utilization in future spintronics. Here, we investigate the response of the single and double AFM skyrmions to   a magnetic field perpendicular to the lattice.    Within our model, as theoretically expected~\cite{rosales2015three,bessarab2019stability,potkina2020skyrmions,aldarawsheh2022emergence}, and in contrast to their FM counterparts, the size of the AFM skyrmions increases with  the external magnetic field, until  its magnitude approaches a critical value ($B_{c}$), after which, the skyrmion deforms into the stripe domain phase. It has been shown that the single and interlinked AFM skyrmions formed with the realistic interactions among Cr atoms  bear high magnetic fields~\cite{aldarawsheh2022emergence}. At the model level, the critical value of the normalised magnetic field ($\mu B_{c}/|J1|$) can be enhanced by increasing the anisotropy magnitude,  as depicted in Fig.~\ref{fig:2}a, for both single  and double AFM skyrmions. In contrast, the DMI lessens the highest magnetic field survived by the AFM solitons,  as shown in Fig.~\ref{fig:2}c. Various formulas have been proposed to describe the impact of DMI  and  anisotropy magnitude on the radius of the FM skyrmions~\cite{rohart2013skyrmion,Zhang2015b,Fert2017,wang2018theory}. Inspired by Ref.~\cite{wang2018theory}, and utilizing the fact that $|J_1| >> D_3,K$,  our results on the dependence of the AFM skyrmion radius $R$ on the anisotropy (Fig.~\ref{fig:2}b) and DMI (Fig.~\ref{fig:2}d) when the external field is switched-off can be fitted with $R_0=a+b\frac{D_3}{K}\left(1+c\frac{D_3^2}{|J_1|K}\right)$,  where a, b and c are fitting parameters.

Upon application of the magnetic field, we found that the form proposed in Ref.~\cite{bessarab2019stability} has to be amended with a linear field-dependent term. After a Taylor expansion in the regime where the field is smaller than the rest of the magnetic interactions, we find \\ $R = a + b\frac{D_3}{K}\left(1+c\frac{D_3^2}{|J_1| K}\right) \left(1+ \alpha \frac{B}{|J_1|}+\beta \frac{B^2}{|J_1|^2} +\gamma \frac{B^3}{|J_1|^3}\right)$, where $\alpha$, $\beta$ and $\gamma$ are additional fitting parameters, grasps reasonably the dependencies reported in Figs.~\ref{fig:2}e-f (with $D_{3}/|J_{1}| =0.03$ and $K/|J_{1}|=0.023$).  

Overall, the magnetic interactions reducing (increasing) the size of the skyrmions, as the magnetic anisotropy (DMI) does, enable an enhanced (reduced) stability with respect to an external magnetic field.

\subsection{AFM skyrmions thermal stability}
\begin{figure}[h!]
	\begin{center}
	\includegraphics[width=1\columnwidth,keepaspectratio]{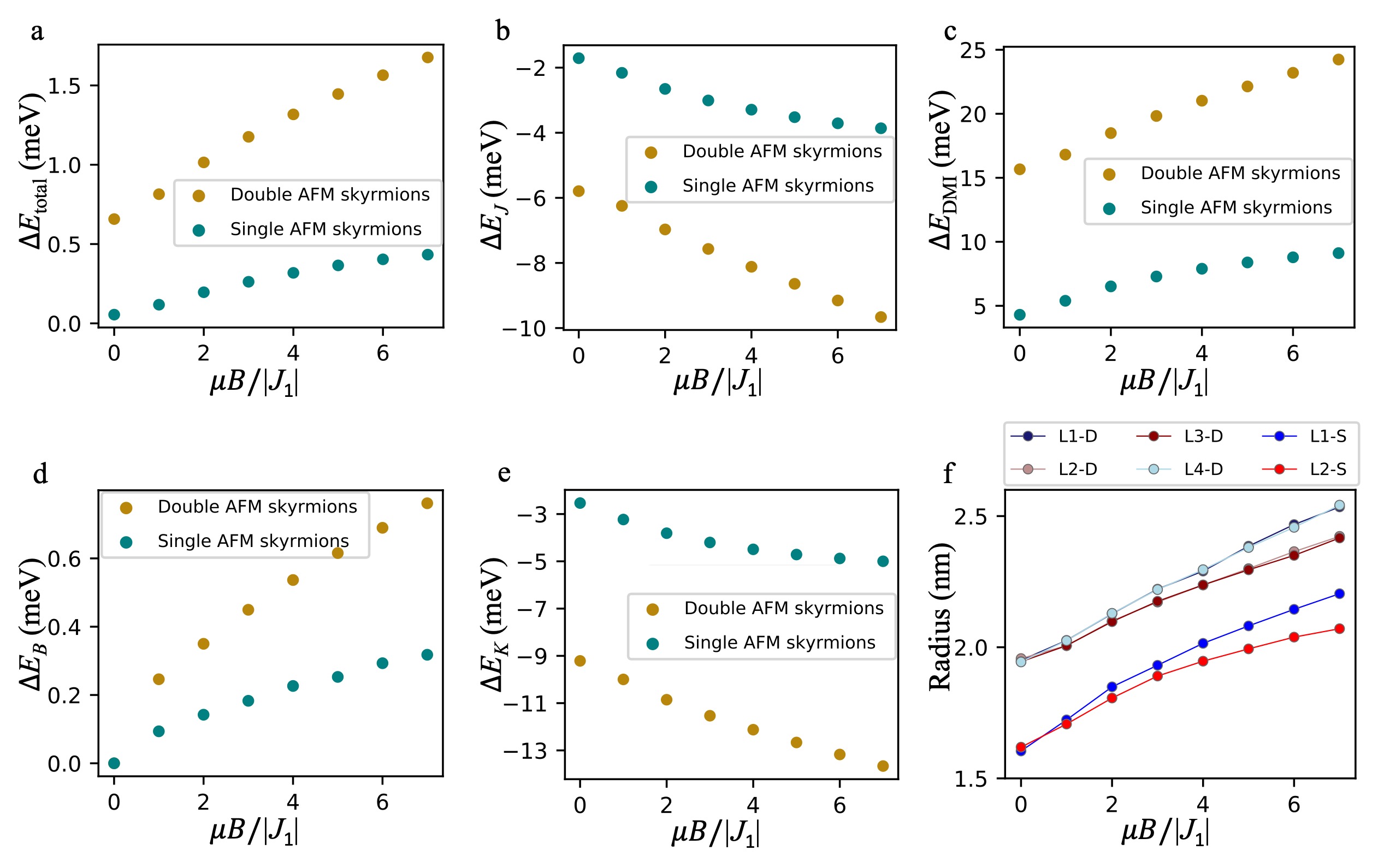} 
	\caption{\textbf{Thermal stability of the single and double AFM skyrmions in the presence of an external magnetic field.} \textbf{a} The total energy barrier for double (golden) and single (green) AFM skyrmions as a function of the normalised value of the external magnetic field. The different contribution to the energy barrier as shown in \textbf{b} exchange, \textbf{c} DMI, \textbf{d} Zeeman and \textbf{e} magnetic anisotropy. \textbf{f} The Radii of the single and double AFM skyrmions are plotted as function of the magnetic field.}
		\label{fig:3}
	\end{center}
\end{figure}

Now we turn to the stability of the AFM skyrmions against thermal fluctuations, by calculating the energy barrier which is needed for the  collapse of the single and double interchained AFM skyrmions into the RW-AFM ground state utilizing the GNEB method~\cite{bessarab2015method,muller2018duplication,muller2019spirit}.
To inspect their stability, we calculate the energy barrier for both single and double interlinked AFM skyrmions assuming  $J_{2}/|J_{1}|$ = -0.2, $J_{3}/|J_{1}|$ = 0.2,  $D_{3}/|J_{1}|$ = 0.03, and  $K/|J_{1}|$ = 0.024. The  barrier is determined by the energy difference between the local minimum magnetic state hosting the AFM skyrmion and its relevant saddle point, which lies on the path of minimum energy connecting the skyrmion configuration to the RW-AFM ground state. In the absence of external magnetic field, the double AFM skyrmions with radius of 1.95 nm, has an energy barrier of 0.67 meV, which translates to $\approx$ 7.8 K, while for the single AFM skyrmion with radius of 1.6 nm, the energy barrier is 0.055 meV ($\approx$ 0.64 K). For both cases, the major key for the stability of the AFM skyrmions  is the DMI which contributes with  $\Delta E_{\mathrm{DMI}}$ =  15.66 meV to the energy barrier of the double AFM skyrmion and 4.33 meV for the single case, while the  anisotropy and exchange interactions  prefer the collapse of the AFM solitons by contributing with $\Delta E_{K}$ = -9.21 meV (-2.53 meV), and  $\Delta E_{J}$ = -5.79 meV (-1.71 meV)  for double (single) AFM skyrmions. Moreover, we addressed another important aspect, the impact of  the magnetic field, by carrying out a systematic study with results illustrated in Fig.~\ref{fig:3}. The thermal stability is obviously enhanced with the magnetic field, which impacts more efficiently the double than the single AFM skyrmion (Fig.~\ref{fig:3}a). For $\mu B/|J_{1}|$ = 1, the energy barrier of the double (single) AFM skyrmions increased to 0.81  meV (0.12 meV) $\approx$ 9.4 K (1.3 K). By increasing the magnetic field, the skyrmions expand (Fig.~\ref{fig:3}f), which in contrast to the DMI and Zeeman  contributions  (Figs.~\ref{fig:3}c-d) is disfavored by those of the exchange  and anisotropy (Figs.~\ref{fig:3}b, e). Snapshots of the various states prospected in defining the energy barriers are presented in Supplementary Figure S4.

\section*{Discussion}
Inspired by our recent findings on the emergence of single and interchained AFM skyrmions on a triangular lattice, we propose here a spin model with the minimum set of magnetic interactions needed to realize such intriguing solitons. They form in a RW-AFM state, which can be decomposed into four sublattices. The exchange interaction within each sublattice, mediating the coupling between the third n.n., is of FM nature along with the associated DMI and out-of-plane anisotropy permit the formation of FM skyrmions within the sublattices. The first n.n. have to be AFM to impose the emergence of AFM skyrmions. We identified the phase diagrams of the latter entities as well as their dependencies on the magnitude of the various magnetic interactions and sensitivity to an external magnetic field. 

The identification of AFM spin-textures requires specific experimental techniques. We expect the recently proposed all-electrical detection based on the tunneling spin-mixing magnetoresistance (TXMR)\cite{crum2015perpendicular, hanneken2015electrical}, with its different possible modes~\cite{lima2022spin} that can be enhanced by the proper implantation of atomic defects~\cite{fernandes2020defect} to be useful for the exploration of AFM states. In this context, the predicted non-collinear Hall effect\cite{bouaziz2021transverse}, the topological spin Hall effect for antiferromagnets~\cite{nakazawa2023topological}  as well as the spin-resolved inelastic electron scattering approaches could be valuable\cite{dos2018spin,dos2020modeling}.
Obviously spin-polarized scanning tunneling microscopy is capable of resolving antiferromagnetic state via atomic resolution\cite{ wortmann2001resolving,schmitt2019indirect}
 while enormous progress has been made with X-ray magnetic microscopy \cite{ juge2022skyrmions} 
 and all-optical relaxometry with a scanning quantum sensor based on a single nitrogen-vacancy (NV) defect in diamond, which were applied for various synthetic AFM textures, among which skyrmions~\cite{finco2021imaging}.
 
 We expect our work to facilitate the search and the identification of single or overlapping AFM skyrmions while contributing to the detailed understanding of their various properties, which is a corner stone in the field of topological antiferromagnetism and its potential use in devices for information technology.



\section{ ACKNOWLEDGEMENTS}
This work was supported by the Federal Ministry of Education and Research of Germany
in the framework of the Palestinian-German Science Bridge (BMBF grant number
01DH16027) and the Deutsche For\-schungs\-gemeinschaft (DFG) through SPP 2137 ``Skyrmionics'' (Project LO 1659/8-1). 
The authors gratefully acknowledge
the computing time granted through JARA on the supercomputer JURECA 
at Forschungszentrum Jülich.


\section{ AUTHOR CONTRIBUTIONS}
S.L. initiated, designed and supervised the project. A.A. performed the simulations with support and supervision from  M.S. A.A., M.S., M.A., and S.L. discussed the results. A.A. and S.L. wrote the manuscript to which all co-authors contributed.
\\

\section{ COMPETING INTERESTS.}
The authors declare no competing interests.

\bibliographystyle{plain}
\bibliography{references}

\end{document}